\documentclass[aps,prd,preprint,superscriptaddress,nofootinbib,preprintnumbers]{revtex4-1}

\usepackage{amsfonts,amsmath,amsthm,amssymb}
\usepackage{mathrsfs}
\usepackage{bm}
\usepackage{color}
\usepackage{float}
\usepackage{epsfig}

\newcommand{\PRE}[1]{{#1}}   

\newcommand{\comment}[1]{}

\newcommand{\zt}{{\tilde z}}

\newcommand{\ba}{\begin{eqnarray}}
\newcommand{\ea}{\end{eqnarray}}
\newcommand{\be}{\begin{equation}}
\newcommand{\ee}{\end{equation}}

\newcommand{\cf}{\textit{cf.} }

\allowdisplaybreaks

\newcommand{\nocontentsline}[3]{}
\newcommand{\tocless}[2]{\bgroup\let\addcontentsline=\nocontentsline#1{#2}\egroup}

\newcommand{\floor}[1]{\lfloor #1 \rfloor}
\newcommand{\ceiling}[1]{\lceil #1 \rceil}

\newcommand{\nwc}{\newcommand}

\nwc{\bea} {\begin{eqnarray}}
\nwc{\eea} {\end{eqnarray}}
\nwc{\nnn} {\nonumber \\[5mm] }

\nwc{\bda} {\bdm\ba{lcl}}
\nwc{\eda} {\ea\edm}

\nwc{\ds}  {\displaystyle}
\nwc{\ra}{\rightarrow}
\nwc{\lra}{\longrightarrow}
\def\lf{\left}\def\ri{\right}

\begin{document}

\preprint{
\hfil
\begin{minipage}[t]{4in}
\begin{flushright}
\vspace*{.3in}
MPP--2013--148
\end{flushright}
\end{minipage}
}

\vspace*{-5cm}
\title{Superstring/Supergravity Mellin Correspondence\\ in Grassmannian Formulation}
\PRE{\vspace*{0.3in}}

\author{Stephan Stieberger}
\affiliation{Max--Planck--Institut f\"ur Physik\\
 Werner--Heisenberg--Institut,
80805 M\"unchen, Germany
\PRE{\vspace*{.1in}}
}
\author{Tomasz R. Taylor}
\affiliation{Department of Physics\\
  Northeastern University, Boston, MA 02115, USA \PRE{\vspace*{.1in}}
}


\PRE{\vspace*{3.5in}}

\begin{abstract}
  \noindent 
We extend the recently established Mellin correspondence of
supergravity and superstring amplitudes to the case 
of arbitrary helicity configurations. The amplitudes are discussed in
the framework of Grassmannian varieties. 
We generalize Hodges' determinant to a function of two sets of
independent coordinates and show that tree-level 
supergravity amplitudes can be obtained by contour integrations of
both sets in separate Grassmannians while in 
superstring theory, one set of coordinates is identified with string
vertex positions at the disk boundary and  
Mellin transformed into generalized hypergeometric functions of Mandelstam invariants.
\end{abstract}

\maketitle

\break\noindent
We have recently established a correspondence between superstring disk amplitudes for the scattering of $N$ gauge bosons and tree-level $N$-graviton supergravity amplitudes, when the  external particles are in the maximally helicity violating (MHV) configurations \cite{Stieberger:2013hza}.
We showed that in both cases, the amplitudes can be obtained from a single generating function -- the Hodges' determinant \cite{Hodges:2012ym} -- in the supergravity case by identifying its arguments with the helicity spinors and in the superstring case by identifying them with string vertex positions on a disk. Integrating over vertex positions amount to  multiple Mellin transforms which convert kinematic variables of supergravity amplitudes into generalized hypergeometric functions of Mandelstam invariants, with the string tension introduced as an energy unit.

In this work, we extend Mellin correspondence to the case of arbitrary helicity configurations. In order to incorporate Yang-Mills building blocks of the amplitudes, we represent them as contour integrals in Grassmannian varieties \cite{ArkaniHamed:2009dn}. We will use  the Veronese map from $G(2,n)\to G(k,n)$ \cite{ArkaniHamed:2009dg}, which is equivalent to the connected prescription in Witten's twistor string theory \cite{Witten:2003nn,Roiban:2004yf}
 and allows a ``particle interpretation" of the integrals. 

We begin by recalling some properties of $N$-graviton MHV amplitudes \cite{Stieberger:2013hza,Hodges:2012ym,sugra,Bern:1998sv,Mason,Feng:2012sy}.
The graviton polarization tensor is defined with reference to  two (gauge fixing) null vectors or respectively, two spinors $x$ and $y$ that can be normalized as $\langle xy\rangle=1$. In the MHV case, Hodges' determinant can be expressed in terms of the Mandelstam invariants $s_{ij}$ and the projective variables
\begin{equation}\sigma_k^1\equiv\langle k x\rangle~,\qquad \sigma_k^2\equiv\langle k y\rangle~,\qquad\qquad k=1,2,\dots N.\end{equation}
In order to write the amplitude, it is convenient to introduce two sets of $\mathbb{CP}^1$ ``position'' coordinates
\begin{equation}z_k=\frac{\sigma_k^1}{\sigma_k^2}~,\qquad
z'_k=\frac{1}{z_k}=\frac{\sigma_k^2}{\sigma_k^1}~.
\end{equation}
Note that
\begin{equation}\label{zmhv1}
z_i-z_j\equiv z_{ij}=\frac{(ij)}{\sigma_i^2\sigma_j^2}~,\qquad z'_i-z'_j\equiv z'_{ij}=\frac{(ji)}{\sigma_i^1\sigma_j^1}\ ,
\end{equation}
where the determinant
\begin{equation}\label{zmhv2}
(ij)\equiv\left|\begin{array}{cc} \sigma^1_i & \sigma^1_j\\
\sigma^2_i  & \sigma^2_j  \end{array}\right| =\langle ij\rangle\ .
\end{equation}
Written in terms of these variables, up to an overall sign, the $N$-graviton MHV amplitude  becomes \cite{Stieberger:2013hza} 
\be A_N^G= \bigg(\prod_{n=1}^N\sigma_n^1\sigma_n^2\bigg)^{-2}\frac{1}{z_{ij}z_{jk}z_{ki}}\frac{1}{z'_{rs}z'_{st}z'_{tr}}\left|\Psi\right|^{rst}_{ijk}\ , \label{mhvh}
\ee
where $\Psi$ is a $N\times N$ ``weighted Laplacian'' matrix \cite{Hodges:2012ym,Feng:2012sy} with the elements
\begin{eqnarray} \psi_{ij}=
\left\{  \begin{array}{ll} \displaystyle \frac{s_{ij}}{z_{ij}z'_{ij}} & \textrm{if}~i\neq j\ ,
\\[3ex]\displaystyle  -\sum_{n\neq i}\frac{s_{in}}{z_{in}z'_{in}}  & \textrm{if}~i= j\ \end{array}\right.\label{psimatrix}\end{eqnarray}
and $\left|\Psi\right|^{rst}_{ijk}$ denotes the minor determinant obtained after deleting three rows $i, j, k$ and three
columns $r, s, t$.

The amplitude is invariant under $\rm GL(2,\!\mathbb{C})$ symmetry reflecting its independence of the choice of reference spinors $(x,y)$. The homogenous coordinates $\sigma_k^{1,2}$ transform as spinor components under the Lorentz subgroup $\rm SL(2,\!\mathbb{C})$, while the position coordinates undergo M\"obius transformations:
\begin{equation}
z_k\to\frac{az_k+b}{cz_k+d}\end{equation}
Note that $z'=1/z$ transform respectively.
The $\rm GL(2,\!\mathbb{C})$ invariance holds for the MHV amplitude evaluated on a momentum-conserving, on-shell kinematic configuration with
\begin{eqnarray}\label{mcon}
\sum_{j=1}^N {}^{{}^{\!\scriptstyle\prime}}~s_{ij}&=&0\ ,\\
\sum_{j=1}^N {}^{{}^{\!\scriptstyle\prime}}~\frac{s_{ij}}{z_{ij}}&=&0\ ,\label{mcon1}
\end{eqnarray}
where the primes over sums denote omission $j\neq i$.

In order to discuss general helicity configurations in supergravity and  in superstring theory, we introduce the density
\be H_N(\sigma,\tilde\sigma)= \bigg(\prod_{n=1}^N\sigma_n^2\tilde\sigma_n^2\bigg)^{-2}\frac{1}{z_{ij}z_{jk}z_{ki}}\frac{1}{\zt_{rs}\zt_{st}\zt_{tr}}\left|X\right|^{rst}_{ijk}\ ,\label{hodges}
\ee
where the $N\times N$ matrix $X$ is defined as
\begin{eqnarray}\label{xmatr} \chi_{ij}=
\left\{  \begin{array}{ll} \displaystyle \frac{s_{ij}}{z_{ij}\zt_{ij}} & \textrm{if}~i\neq j\ ,
\\[3ex]\displaystyle  -\sum_{n\neq i}\frac{s_{in}}{z_{in}\zt_{in}}  & \textrm{if}~i= j\ .\end{array}\right.\label{phimatrix}\end{eqnarray}
The above expression has the same form as the MHV amplitude (\ref{mhvh}), but now it is considered as a function {\em two independent} sets of coordinates:
\be
\sigma^{1,2}_k:~z_k=\frac{\sigma_k^1}{\sigma_k^2}~,\qquad\qquad\tilde\sigma^{1,2}_k:~\zt_k=\frac{\tilde\sigma_k^1}{\tilde\sigma_k^2}~.
\ee
Here again,
\begin{equation}
 z_{ij}=\frac{(ij)}{\sigma_i^2\sigma_j^2}~,\qquad \zt_{ij}=\frac{(\tilde{\imath}\tilde{\jmath})}{\tilde\sigma_i^2\tilde\sigma_j^2}\  ,
\end{equation}
however, {\em unlike} in Eqs. (\ref{zmhv1},\ref{zmhv2}), these coordinates are {\em not\/} yet related to helicity spinors or other kinematic variables. 

Both superstring  and supergravity amplitudes contain Yang-Mills building blocks.  The Kawai-Lewellen-Tye (KLT) construction \cite{Kawai:1985xq} of supergravity amplitudes involves ``gluing" two Yang Mills amplitudes with appropriate kernels while the open superstring amplitudes can be obtained by gluing Yang-Mills
with certain world-sheet integrals \cite{string,Mafra,Stieberger:2012rq}. Following this path, we will obtain (one) Yang-Mills block of N$^{k-2}$MHV amplitudes by a contour integration over the Grassmannian variety $G(k,\! N)$, with the variables $\tilde\sigma$ (and $\zt$) identified as the base coordinates of the Veronese map $(\mathbb{C}^2)^N\!/{\rm GL(2) }\to G(k,\! N)$ \cite{ArkaniHamed:2009dg}. The  remaining $\sigma$ (and $z$) variables will be associated to  Grassmannian coordinates of the second Yang-Mills block or to string vertex positions. In either case, we need to show that the density $H_N$ (\ref{hodges}) is invariant under a larger $\rm SL(2)\times \widetilde{\rm SL}(2)$ symmetry group acting on $\sigma$ and $\tilde\sigma$. 

The Grassmannian integration over $\tilde\sigma$ involves delta function constraints relating these coordinates to kinematic variables in a way respecting momentum conservation \cite{ArkaniHamed:2009dn}. Actually, an explicit $\delta(\sum\! p)$  factor can be always extracted by appropriate change of  integration variables. These kinematic relations are also responsible for the Bern-Carrasco-Johansson (BCJ) relations \cite{Bern:2008qj}  between partial amplitudes. In this context, Cachazo \cite{Cachazo:2012uq} showed that:
\be
\sum_{j=1}^N {}^{{}^{\!\scriptstyle\prime}}~ s_{ij}\frac{(\tilde n\tilde \jmath)}{(\tilde \imath\tilde \jmath)}=0\ ,
\ee
which, together with Eq. (\ref{mcon}), imply
\be
\sum_{j=1}^N {}^{{}^{\!\scriptstyle\prime}}~\frac{s_{ij}}{\zt_{ij}}=0\ ,\label{zconst}
\ee
thus generalizing Eq. (\ref{mcon1}) to arbitrary helicity configurations. This relation can be used for our purpose, to show  that under $\rm SL(2)$, the
$X$ matrix  elements (\ref{xmatr}) transform as 
\be \chi_{ij}\to (cz_i+d)(cz_j+d)\,\chi_{ij}\ ,
\ee
thus in Eq. (\ref{hodges}), the corresponding factor from the determinant  is canceled by the factor supplied by $(z_{ij}z_{jk}z_{ki})^{-1}(\prod_{n=1}^N\sigma_n^2)^{-2}$. Hence $H_N$ is invariant under $\rm SL(2)$ transformations.

In order to demonstrate $\widetilde{\rm SL}(2)$-invariance, we need to differentiate between supergravity and superstring cases.
In supergravity, $\sigma$ and $z$ variables will be associated to their own Grassmannian variety, hence they will be constrained in the same way as $\zt$ in
Eq. (\ref{zconst}), and $\widetilde{\rm SL}(2)$-invariance follows by the same argument.
{}In superstring disk amplitudes, the $\rm SL(2)$-invariant density $H_N$ will be weighted by Koba-Nielsen factors and integrated over $N$ (ordered) $z$ variables now identified with string vertex positions on $\mathbb{P}^1$. The corresponding string ``formfactor'' integrals are constrained by the vanishing integral of the total derivative
\be\frac{\partial}{\partial z_i}\prod_{1\leq k<l\leq N}|z_{kl}|^{s_{kl}}=\sum_{j=1}^N {}^{{}^{\!\scriptstyle\prime}}~\frac{s_{ij}}{z_{ij}}\prod_{1\leq k<l\leq N}|z_{kl}|^{s_{kl}}\ .
\ee
Integrations by parts resulting in such vanishing total derivative terms were used in Ref. \cite{Broedel:2013tta}  to derive BCJ relations for $N$-point disk integrals. The bottom line is that the relation 
\be
\sum_{j=1}^N {}^{{}^{\!\scriptstyle\prime}}~\frac{s_{ij}}{z_{ij}}=0\label{bcjz}
\ee
can be used in both supergravity and superstring amplitudes to show that up to total derivatives, the density $H_N$ is invariant under $\widetilde{\rm SL}(2)$ transformations. Furthermore, BCJ relations can be used on both ${\rm SL}(2)$ and $\widetilde{\rm SL}(2)$ sides.

Now we can use  $\rm SL(2)\times \widetilde{\rm SL}(2)$ invariance to make connections with the KLT formula 
\cite{Kawai:1985xq} for gravity amplitudes and with the general formula for the disk amplitudes \cite{Mafra,Broedel:2013tta}. Although this can be done for an arbitrary set of $i,j,k,r,s,t$ indices in Eq. (\ref{hodges}),
in order to streamline the argument, we
choose $i=r=1, ~j=s=N{-1}, ~k=t=N$. Then, according to the matrix--tree theorem, the determinant is given by the sum of all forests consisting of three trees rooted at $1$, $N{-}1$ and $N$, with a combined number of $N{-}3$ edges,  each of them bringing a $\chi_{ij}$ factor \cite{Feng:2012sy}. 
At this point, we go to a $\rm SL(2)$ reference frame with $\sigma^2_{N}=0$  $(z_N\to\infty)$ and to a $\widetilde{\rm SL}(2)$  reference  frame with $\tilde\sigma^2_{N{-}1}=0$  $(\zt_{N{-}1}\to\infty)$.
In this way, we are left with single trees only, rooted at $1$, {\em
  i.e}.\ all trees with $N{-}2$ vertices different from $N{-}1$ and
$N$. Next, we partial fraction $\zt_{ij}$ denominators on the
$\widetilde{\rm SL}(2)$ side, as described in Ref. \cite{Stieberger:2013hza}, use BCJ relations on the $\rm SL(2)$ side and finally restore $\rm SL(2)\times \widetilde{\rm SL}(2)$
invariance by returning to a general frame. All this seems rather involved, so it is worth illustrating on some simple examples.

{}For $N=4$, we begin with
	\be H_4(\sigma,\tilde\sigma)=\frac{1}{(13)(14)(34)}\frac{1}{(\tilde 1\tilde 3)(\tilde 1\tilde 4)(\tilde 3\tilde 4)}\frac{1}{(\sigma_2^2)^2(\tilde\sigma_2^2)^2}\frac{s_{12}}{z_{12}\zt_{12}}
\ee
which, in the reference frame with $\sigma_4^2=\tilde\sigma_3^2=0$, becomes
\be H_4(\sigma,\tilde\sigma)=\frac{1}{\tilde \sigma_2^2\tilde\sigma_3^1(\tilde 1\tilde 4)(\tilde 3\tilde 4)}\frac{1}{ \sigma_2^2\sigma_4^1(13)(34)}\frac{s_{12}}{(12)(\tilde 1\tilde 2)}\ .\ee
After reverting to a general reference frame,
\be H_4(\sigma,\tilde\sigma)=\frac{1}{(\tilde 1\tilde 2)(\tilde 2\tilde 3)(\tilde 3\tilde 4)(\tilde 4\tilde 1)}~s_{12}~\frac{1}{(12)(24)(43)(31)}\ .\label{hfour}\ee
Starting from $N=5$, partial fractioning and BCJ relations become very helpful. After partial fractioning, we obtain:
\be H_5(\sigma,\tilde\sigma)=\frac{1}{(14)(15)(45)}\frac{1}{(\tilde 1\tilde 4)(\tilde 1\tilde 5)(\tilde 4\tilde 5)}\frac{1}{(\sigma_2^2\sigma_3^2)^2(\tilde\sigma_2^2\tilde\sigma_3^2)^2}\frac{1}{\zt_{12}\zt_{23}}\frac{s_{12}}{z_{12}}
\bigg(\frac{s_{13}}{z_{13}}+\frac{s_{23}}{z_{23}}\bigg) ~+~ (2\leftrightarrow 3)\ .
\ee
BCJ relations are implemented by applying Eq. (\ref{bcjz}) to the bracket on the r.h.s.\ which, in the reference frame of $z_5\to\infty$, becomes:
\be
\frac{s_{13}}{z_{13}}+\frac{s_{23}}{z_{23}}=\frac{s_{34}}{z_{34}}\ .\ee
Staying in this frame, we obtain:
\be H_5(\sigma,\tilde\sigma)=\frac{1}{\tilde \sigma_3^2\tilde\sigma_4^1(\tilde 1\tilde 5)(\tilde 4\tilde 5)(\tilde 1\tilde 2)(\tilde 2\tilde 3)}\,\frac{s_{12}s_{34}}{\sigma_2^2\sigma_5^1\sigma_3^2\sigma_5^1(14)(12)(34)}
~+~ (2\leftrightarrow 3)\ .
\ee
After reverting to a general reference frame, we obtain
\be H_5(\sigma,\tilde\sigma)= \widetilde M(1,2,3,4,5)~ s_{12}s_{34}~ M(2,1,4,3,5)~+~ (2\leftrightarrow 3)\ ,\label{hfive}
\ee
with the definitions
\bea M(i_1,i_2,\dots,i_N)&=&\frac{1}{(i_1i_2)(i_2i_3)\cdots(i_Ni_1)}\ ,\nonumber \\
\widetilde M(i_1,i_2,\dots,i_N)&=&\frac{1}{(\tilde\imath_1\tilde\imath_2)(\tilde\imath_2\tilde\imath_3)\cdots(\tilde\imath_N\tilde\imath_1)}\ ,
\eea
which become useful for higher $N$.

The $N=4$ and $N=5$ examples should make it clear how to proceed to higher $N$. 
The result can be written in many ways, the shortest involving
$(N-3)!\times (\floor{\tfrac{N}{2}}-1)!\times (\ceiling{\tfrac{N}{2}}-2)!$ terms\footnote{$\floor{x}$ is the integer part of $x$, while  $\ceiling{x}$ gives the smallest integer greater than or equal to $x$.} being:\vskip 5mm
\begin{eqnarray} H_N(\sigma,\tilde\sigma)&=&\sum_{\pi\in S_{N{-}3}}\! \widetilde M(1,\pi(2,3,\dots,N{-}2),N{-}1,N)
\nonumber\\[1mm] && \times\!\!\sum_{\alpha\in S_{\floor{N/2}{-}1}}\!\!
\widetilde S(\alpha\circ\pi(2,\ldots,\floor{N/2}))\sum_{ \beta\in S_{\ceiling{N/2}{-}2}}\!\!
 S(\beta\circ\pi(\floor{N/2}{+}1,\dots N{-}2))\nobreak
\nonumber\\[2.5mm] &&\nobreak
\times~ M(\alpha\circ\pi(2,\dots,\floor{N/2}),1,N-1, \beta\circ\pi(\floor{N/2}{+}1 ,\ldots,N{-}2), N       )\ ,\label{hsug}
\end{eqnarray}
where $\pi$, $\alpha$ and $\beta$ denote permutations of the respective numbers of $N-3$, $\floor{N/2}-1$ and
$\ceiling{N/2}-2$ elements. Here,
\bea
\widetilde S(i_1,\ldots,i_j)&=&s_{1i_j}\ \prod_{m=1}^{j-1}\lf(s_{1i_m}+\sum_{k=m+1}^j
\theta(i_m,i_k)\ri)\ ,\nonumber\\
S(i_1,\ldots,i_l)&=&s_{i_1N{-}1}\ \prod_{m=2}^{l}\lf(s_{i_mN{-}1}+\sum_{k=1}^{m-1}\theta(i_k,i_m)\ri)\ ,
\eea
with
\be
\theta(i,j)=\left\{
\begin{array}{ll}s_{ij}\ ,& i>j\ ,\\
0\ ,& i<j\ ,
\end{array}\right.
\ee
are the elements of the KLT momentum kernel 
\cite{Kawai:1985xq,Bern:1998sv,Bohr}. We conclude that the generalized Hodges' determinant density $H_N$ of Eq. (\ref{hodges}) does indeed describe two Yang-Mills factors ``glued" by KLT kernels. We will show below that it yields correct supergravity amplitudes and agrees with the general formula for open superstring amplitudes, when integrated with appropriate delta function constraints and Koba-Nielesen factors.

We are ready to apply the generalized Hodges' determinant as a pivot for connecting supergravity and superstring amplitudes in the Grassmannian framework.
The tree level, N$^{k-2}$MHV supergravity amplitude is given by
\begin{eqnarray} A_{k,N}^G &=&
\frac{1}{\rm vol(GL(2))vol(\widetilde{\rm GL}(2))}\int d^2\sigma_1\cdots d^2\sigma_{N}\int d^2\tilde\sigma_1\cdots d^2\tilde\sigma_{N}\, H_N(\sigma,\tilde\sigma)\nonumber\\[1mm] &&
\times\,\prod_{\alpha=1}^{k}\delta^{4|4}\big(\sum_{i=1}^NC_{\alpha i}^V[\sigma]{\cal W}_i(\eta)\big)
\prod_{\tilde\alpha=1}^{k}\delta^{4|4}\big(\sum_{i=1}^NC_{\tilde\alpha i}^V[\tilde\sigma]{\cal W}_i(\tilde\eta)\big)
\ ,\label{ag}
\end{eqnarray}
where $C^V$ is the Veronese map \cite{ArkaniHamed:2009dg}. The
kinematic data are specified by the (dual) ${\cal N}=4$ supertwistors
${\cal W}$. Note the presence of two sets of anticommuting variables,
$\eta$ and $\tilde\eta$ which are necessary for the enhancement to
${\cal N}=8$ supersymmetry. Eq. (\ref{hsug}) guarantees agreement with
the KLT formula however it is easy to see that Eq. (\ref{ag}) overshoots by an extra  $\delta(\sum\! p)$  factor - thus a pedantic reader should divide the r.h.s.\ by $\delta^4(0)$. We should note that the Grassmannian formulation of supergravity amplitudes and their twistor string origin have been considered before in Refs.\cite{Cachazo:2012da,twist};  in particular, Ref. \cite{Cachazo:2012da} makes a similar connection between Hodges' determinant and KLT formula.

The partial N$^{k-2}$MHV open superstring amplitude is given, at the disk level, by
\begin{eqnarray} A_{k,N}^S &=&
\frac{1}{\rm vol(SL(2))vol(\widetilde{\rm GL}(2))}\int_{D\subset (\mathbb{P}^1)^N}\!\! d^2\sigma_1\cdots d^2\sigma_{N}\int d^2\tilde\sigma_1\cdots d^2\tilde\sigma_{N}\, H_N(\sigma,\tilde\sigma)\nonumber\\[1mm] &&
\times\prod_{1\leq i<j\leq N}|(ij)|^{s_{ij}}
\prod_{\tilde\alpha=1}^{k}\delta^{4|4}\big(\sum_{i=1}^NC_{\tilde\alpha i}^V[\tilde\sigma]{\cal W}_i(\tilde\eta)\big)
\ ,\label{as}
\end{eqnarray}
where the domain of $\sigma$-integration $D\subset (\mathbb{P}^1)^N$ is determined by color ordering.
Eq. (\ref{hsug}) guarantees agreement with the general formula for disk amplitudes \cite{Mafra}, as recently recast in the KLT form in Ref. \cite{Broedel:2013tta}. Note that the string integral, including the Koba-Nielsen factor $\prod_{1\leq i<j\leq N}|(ij)|^{s_{ij}}\sim\prod_{1\leq i<j\leq N}|z_{ij}|^{s_{ij}}$, is fully ${\rm GL}(2)$ invariant, but the integration domain is restricted to  the disk boundary where string vertex positions are specified by $z$ coordinates\footnote{Here, one also overshoots by an extra  $\delta(\sum\! p)$  factor which in this case originates from the world-sheet correlator giving rise to the Koba-Nielsen factor.}. When integrating on $\mathbb{P}^1$, we can set $\sigma^2_i=1$, so that $(ij)= z_{ij}$ and replace $d^2\sigma_i\to dz_i$. In Ref. \cite{Stieberger:2013hza}, we  argued that integrations with Koba-Nielsen factor amount to Mellin transforms with the measure
\be
\int dM_N=\frac{1}{\rm vol(SL(2))}\ \int_{D\subset (\mathbb{P}^1)^N}\!\! 
dz_1\ldots dz_{N}\ \prod_{1\leq i<j\leq N}|z_{ij}|^{s_{ij}}\ ,
\ee
which are multi--dimensional Mellin transforms from the string world--sheet boundary to the Mellin space of Mandelstam invariants encoded in  generalized hypergeometric functions. 
Thus Eqs. (\ref{ag}) and (\ref{as})  extend the supergravity/superstring Mellin correspondence from the MHV case considered  in Ref. \cite{Stieberger:2013hza} to all helicity configurations.

It is clear that the generalized Hodges' determinant (\ref{hsug}) plays a central role in superstring/supergravity correspondence. 
This may indicate the existence of yet another, perhaps more fundamental theory. On the other hand, one could try to formulate Mellin correspondence without referring to this determinant. 
In principle, in order to construct string amplitudes, one could start from field-theoretical amplitudes or some motivic objects (\cf \cite{Motivic}) taken as functions of kinematic variables and  consider transformations with respect to some dual variables. This may be difficult though because there are many, often highly non-trivial but equivalent, ways of writing the amplitudes.

Field theory appears in the $\alpha'\to 0$ limit of string theory. Going the other way, directly from field theory to full-fledged string theory by means of integral transforms is harder, but it may be possible with some better understanding of the amplitudes.

\vskip 1.5cm

\noindent
\textbf{Acknowledgements}
\\[5mm]   
We thank Nima Arkani-Hamed,  Sergio Ferrara and David Skinner
for useful discussions and correspondence.  Furthermore, we are grateful to the Theory Division of CERN for hospitality and financial support when
finalizing this work. T.R.T.\ is grateful to Max-Planck-Institut f\"ur Physik in M\"unchen for kind hospitality.
This material is based in part upon work supported by the National Science Foundation under Grant No.\ PHY-0757959.  Any
opinions, findings, and conclusions or recommendations expressed in
this material are those of the authors and do not necessarily reflect
the views of the National Science Foundation.

\end{document}